# Landfalling tropical cyclones perturb Earth's rotation

Yaxuan Yang[1], Lei Liu[1], Zhouyu Zhang[1,2], Yongjie Liu[1], Qingyuan Wang[1*], and Lin Li[1*]

**Affiliations:**

[1]Failure Mechanics and Engineering Disaster Prevention Key Laboratory of Sichuan Province, Sichuan University, Chengdu 610207, China

[2]Department of Mechanical Engineering, University of Michigan, Ann Arbor, MI, 48109, United States

*Corresponding author. Email: wangqy@scu.edu.cn; lin.li@scu.edu.cn;

**Natural disasters that devastate human societies can also alter Earth's rotation. The 2011 Mw 9.1 Tōhoku earthquake, one of the most powerful earthquakes ever recorded, shortened the day by 1.80 μs. Tropical cyclones are the costliest weather disasters, yet whether they can alter Earth's rotation remains unknown. Here we use atmospheric reanalysis to estimate, and numerical simulations to confirm, that a landfalling tropical cyclone can transiently alter the length of day by amounts rivalling those by great earthquakes. For example, Hurricanes Harvey (2017) and Katrina (2005) lengthened the day by 1.87 μs and 2.29 μs, respectively. As the climate warms, tropical cyclones may become an increasingly important source of variability in Earth's rotation.**

## Introduction

Extreme natural events that devastate human societies often set vast masses of rock, water and air in motion. Conservation of angular momentum couples these motions to the spin of the solid Earth, allowing such events to alter the length of day (LOD)[1,2,3]. These natural events are only the most dramatic members of a broader family of processes that continually perturb Earth's rotation, and these processes fall into two fundamentally different classes. The first is the torque-driven exchange of angular momentum between the solid Earth and other reservoirs, including the atmosphere[2,3], oceans[4], liquid core[5] and Moon[6]. The second is internal mass redistribution, which changes the moment of inertia of the solid Earth, as in great earthquakes[7,8] and glacial isostatic adjustment[9]. The 2011 Mw 9.1 Tōhoku earthquake[10], for example, shortened the day by 1.80 μs[11]. Earthquakes have thus become the canonical example of a natural disaster altering Earth's rotation. Yet within the first class, one of the most violent and concentrated angular-momentum exchanges on Earth has received comparatively little attention: the landfall of a tropical cyclone.

A tropical cyclone is a compact reservoir of angular momentum gathered from a far larger volume of atmosphere[12,13]. As it intensifies over the warm ocean, air converges from as far as ~1,000 kilometres away, and the planet's weak background rotation is concentrated into a core several hundred kilometres wide, where the local angular velocity, $v_\theta/r$, can exceed Earth's rotation rate by up to two orders of magnitude. As a tropical cyclone makes landfall, its energy supply from the ocean is cut off, and enhanced drag over the rough land surface exerts a strong frictional torque that rapidly damps the circulation (timescale of ~1 day)[14,15]. This rapid coupling to the solid Earth makes landfall the natural stage at which to seek the strongest signature in LOD. It also sets up a tempting first picture: a landfalling tropical cyclone, rotating in the same sense as Earth in either hemisphere, behaves like a flywheel being braked, transferring prograde angular momentum to the solid Earth and therefore always shortening the day. An earlier idealised estimate based on this picture predicted a LOD shortening of about 0.014 μs[16], two orders of magnitude smaller than that of a great earthquake. Whether real tropical cyclones alter Earth's rotation in this way, and if so by how much, is the question we take up in our study.

## Results

### Reanalysis-based estimates of storm-induced perturbations

We analyse twelve high-impact landfalling tropical cyclones in the North Atlantic, where the observational record is most reliable[17]. Ten are the basin's costliest[18] such storms once we exclude those that underwent extratropical transition soon after landfall, crossed only narrow landmasses or remained over land for less than a day; to these we add Mitch (1998) and Isidore (2002) for their contrasting landfall directions (Methods and Extended Data Table 1). For each storm, we quantify the angular-momentum exchange with the solid Earth using ERA5, the fifth-generation reanalysis of the European Centre for Medium-Range Weather Forecasts, which is regarded as the closest available estimate of the real atmosphere[19]. In a storm-following frame, we centre a 500-km-radius disc on the storm at each time step and compute the axial torque $\Gamma$ on the solid Earth by integrating the zonal (east–west) component of the surface stress over land within the disc, multiplying each surface element by its moment arm, the distance from Earth's rotation axis (Methods). This 500-km radius is comparable to the climatological outer extent of tropical cyclones, whose median radius of vanishing wind is 423 km[20]; because surface stress scales quadratically with wind speed, the disc captures essentially all of the storm-induced torque while admitting little contamination from the ambient flow. We integrate Γ from the moment the disc first overlaps land until HURDAT2[17] indicates that storm has either weakened below tropical-storm intensity (17.5 m $s^{-1}$) or became extratropical. Then we convert the accumulated angular-momentum transfer, $\Delta L = \int \Gamma dt$, into an equivalent change in the length of day, ΔLOD.

Remarkably, eight of the twelve storms exceed the Mw 9.1 Tōhoku earthquake[11] in |ΔLOD| (Fig. 1a). The largest response was produced by Hurricane Mitch (1998), which shortened the day by 3.43 μs. Hurricane Katrina (2005) and Hurricane Harvey (2017), two of the costliest storms, lengthened it by 2.29 μs and 1.87 μs, respectively. Of all recorded earthquakes, only the very largest, the 1960 Chile, 1964 Alaska and 2004 Sumatra earthquakes, produced larger |ΔLOD|[7,11,21]. The storm-induced ΔLOD is not only much larger than the idealised flywheel estimates but also of either sign. This directly contradicts the braked-flywheel picture: had landfall merely drained the storm's angular momentum into the solid Earth, every storm would have shortened the day (ΔLOD < 0). Instead, ten of the twelve lengthened it (ΔLOD > 0). What is missing in the flywheel picture?

A first clue lies in the storms' tracks: the sign of ΔLOD depends systematically on the direction in which a storm approaches land (hereafter "landfall direction"). Hurricanes Mitch and Katrina, which approached land in nearly opposite directions, exemplify this dependence (Fig. 1b,c). Mitch moved southward into Central America and shortened the day; Katrina moved northward across the U.S. Gulf Coast and lengthened the day.

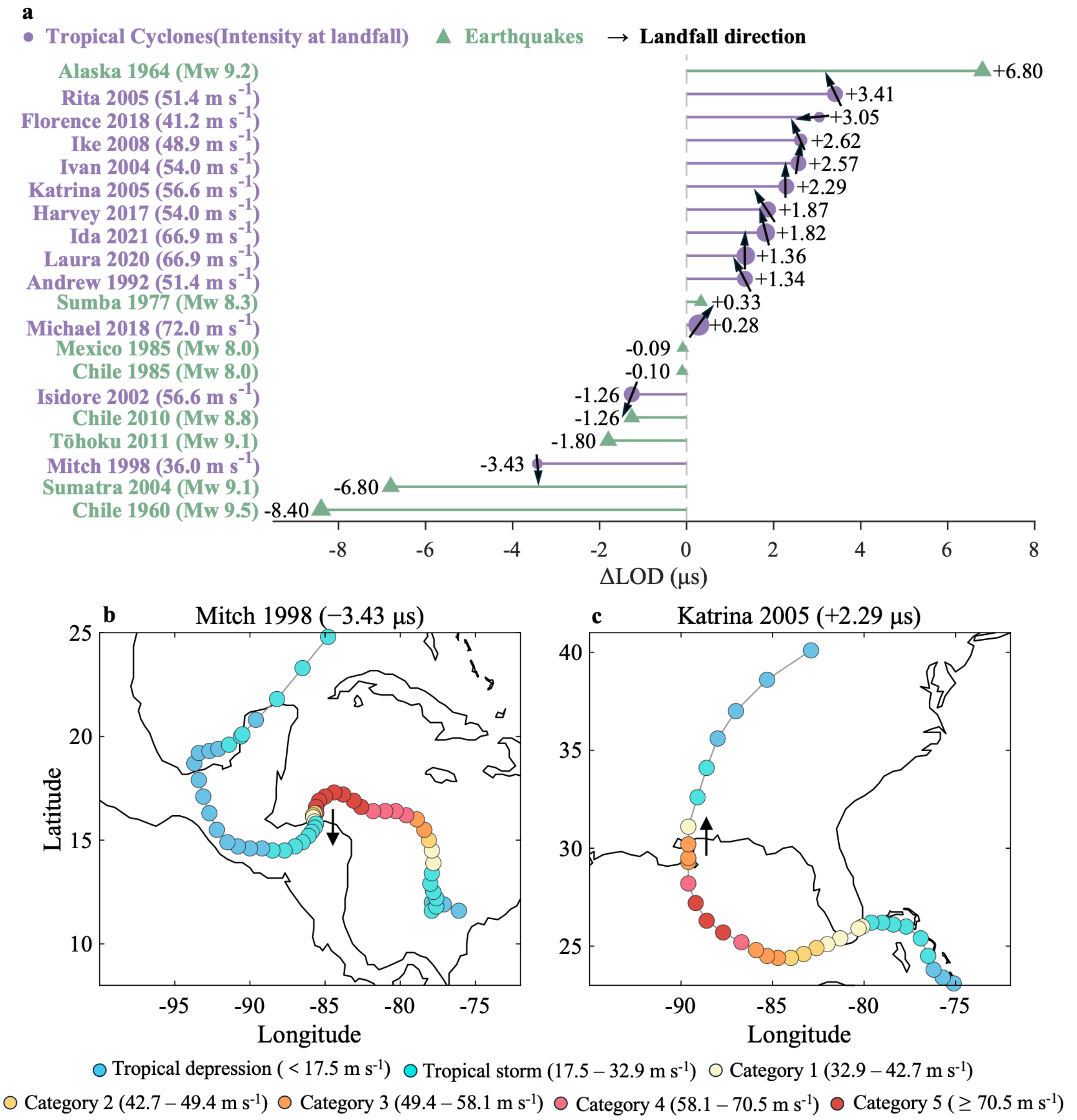

**Figure 1 | Storm-induced ΔLOD and two contrasting landfall tracks.**

**a,** ΔLOD for 12 landfalling tropical cyclones (purple circles) and 8 great earthquakes (Extended Data Table 2). Marker size scales with storm intensity at landfall or earthquake magnitude. Arrows indicate the direction of storm motion at landfall (landfall direction). **b,c,** Best tracks of Hurricanes Mitch (1998; b) and Katrina (2005; c), which approached land in nearly opposite directions and produced ΔLOD of opposite sign (−3.43 μs and +2.29 μs); colours denote storm intensity (Saffir–Simpson scale) and arrows mark the landfall direction.

**Idealised numerical experiments**

To isolate the mechanism underlying this directional dependence, we perform an idealised landfall simulation with Cloud Model 1[22,23] , a three-dimensional, non-hydrostatic computational model that has been widely used to study tropical cyclones. The simulation is conducted with a constant Coriolis parameter $f = 5.0 \times 10^{-5}$ s$^{-1}$ (corresponding to the latitude at 20° N), in a 6,000 km × 3,000 km domain divided by a straight coastline into equal ocean and land halves. We first develop a mature tropical cyclone over ocean, and then steer it across the coastline by an imposed flow of 5 m s$^{-1}$, a translation speed typical of observed tropical cyclones[24,25]. The storm's intensity at landfall (when its centre crosses the coastline) is 57.3 m s$^{-1}$. Over land, we set the surface enthalpy exchange coefficient $C_E$ to zero, suppressing surface moisture and heat fluxes, and the drag coefficient $C_D$ to 0.005, implying a rougher surface than the ocean[26] (see Methods and Extended Data Table 3).

Because the governing equations use a constant Coriolis parameter and contain no preferred horizontal direction, the simulated landfall can be rotated without altering its dynamics. We exploit this property by diagnostically rotating the entire simulated system, including the track, coastline, wind field and surface stress field, about the fixed landfall point. These rotations preserve the storm's evolution, translation speed, and angle of incidence to the coastline, while changing only its landfall direction. For each orientation, we map the rotated fields onto geographic coordinates centred on the fixed landfall point at 20°N and calculate the axial torque by integrating the zonal component of the atmospheric stress on land over the storm-following disc, multiplying each surface element by its moment arm, $R\cos\varphi$, where $R$ is Earth's radius and $\varphi$ is latitude. We then integrate the resulting torque over land, from the moment the disc first overlaps land until the storm's surface wind weakens below the tropical-storm threshold of 17.5 m s$^{-1}$, yielding the angular momentum transferred to the solid Earth and the corresponding ΔLOD. This procedure isolates the effect of geographic approach direction from differences in storm structure, intensity or decay rate. Simply changing this direction, with everything else held fixed, is enough to reverse the sign of ΔLOD (Fig. 2). ΔLOD varies nearly sinusoidally with landfall direction, reaching +0.72 μs near 300° and −0.75 μs near 120°, and crossing zero near 30° and 210°. (The amplitude is smaller than the observed values, which is mainly due to the smaller radius of maximum wind and smoother land surface in the simulation.) Indeed, this directional dependence is consistent with the observed storms. The largest responses come from storms that moved northwestward or southeastward at landfall; Rita (2005), for example, moved northwestward and produced the second largest response (+3.41 μs). Conversely, Michael (2018), despite being the most intense of the twelve storms at landfall, moved northeastward, and produced the smallest response (+0.28 μs). We next ask what sets this directional dependence.

Consider first a storm heading north (Fig. 3a,c). As it moves onto land, its northern flank, where the winds blow westward, is the first to reach land. This westward wind exerts a torque on the solid Earth that opposes its rotation and lengthens the day. Once the storm is fully over land, both flanks act on the surface, and the net torque is set by the difference between them. This contribution, however, is much smaller than the initial one-sided torque: the winds have decayed by then, and for a symmetric wind field the torques from the two flanks nearly cancel, differing only through their slightly different moment arms. The sign of ΔLOD is thus set by the zonal wind of whichever flank reaches land first. A southward-moving storm mirrors this sequence and reverses the sign of ΔLOD. This mechanism explains the observed contrast between the northward-moving Katrina, which lengthened the day, and the southward-moving Mitch, which shortened it (Fig. 1b,c).

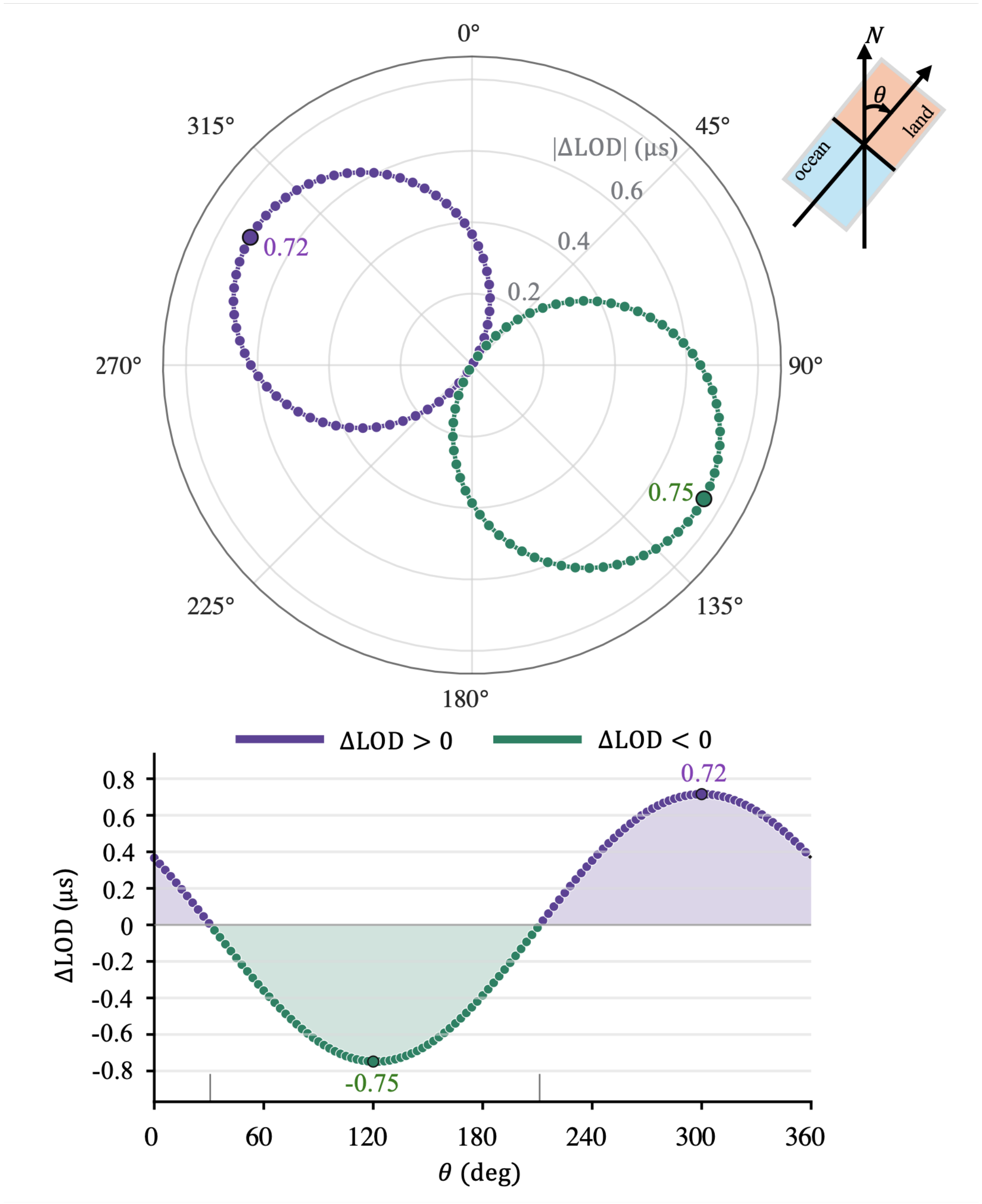

**Figure 2 | Landfall direction sets the sign and magnitude of storm-induced ΔLOD.**

**a,** Landfall direction $\theta$, measured clockwise from north, with $\theta = 0°$ corresponding to a south-to-north landfall (inset). Radius gives |ΔLOD| (μs); colour indicates the sign of ΔLOD (purple, positive ΔLOD; green, negative ΔLOD). **b,** The same data plotted as signed ΔLOD against $\theta$, showing the predominantly sinusoidal dependence. ΔLOD reaches a maximum of +0.72 μs at $\theta = 300°$ and a minimum of −0.75 μs at $\theta = 120°$, reversing sign near $\theta = 30°$ and $\theta = 210°$. Enlarged markers in both panels denote the extrema.

Now consider a storm heading west (Fig. 3b,d). Its western flank reaches land first, but unlike the northern flank of a northward-moving storm, this flank carries zonal winds of both signs, which would cancel for a symmetric storm. The sign of ΔLOD is therefore set by whatever breaks the north–south symmetry, and the steering flow does so. The westward steering flow adds to the westward winds north of the centre and subtracts from the eastward winds south of it, the well-known strengthening of winds to the right of the direction of motion[27]. Because surface stress grows as the square of the wind speed, this modest asymmetry in wind becomes a large asymmetry in stress. So, the net torque opposes Earth's rotation, and the day lengthens, as found in Fig. 2. An eastward-moving storm mirrors this sequence and shortens the day. Latitude adds a second, opposing asymmetry: the southern half lies farther from the Earth rotation axis and acts with a longer moment arm, favouring a shorter day. Across the storm, however, the moment arm differs by a few percent, roughly two orders of magnitude less than the steering-induced stress asymmetry (Extended Data Fig. 1 and Methods), so it cannot reverse the sign.

For other landfall directions, the net impulse imparted to land by surface stress combines a cross-track component set by the leading flank with an along-track component set by the steering-induced asymmetry. Their contributions to ΔLOD reinforce each other for northwestward and southeastward landfalls and oppose each other for northeastward and southwestward landfalls. This yields a nearly sinusoidal dependence on landfall direction in Fig. 2b. We test this interpretation using a single-layer, purely mechanical model, in which a Rankine vortex is carried by a uniform steering flow and decays after landfall at a prescribed rate. The model qualitatively reproduces this angular dependence (Extended Data Fig. 1a) and shows that the small departure from this sinusoidal dependence arises from the latitudinal variation in the moment arm (Extended Data Fig. 1b).

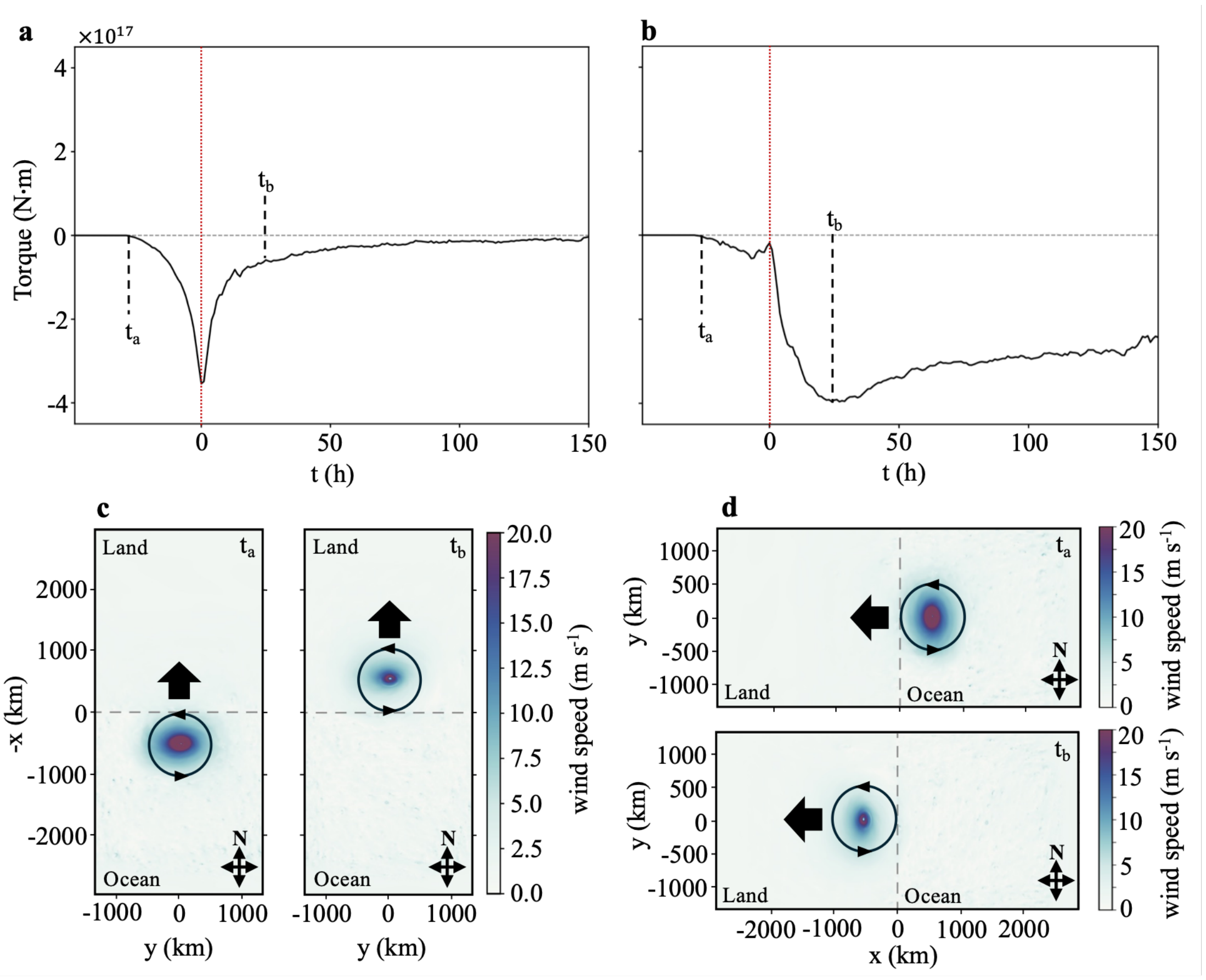

**Figure 3 | Time evolution of the axial torque for northward and westward landfalls.** **a,b,** Axial torque $\Gamma$ exerted on the solid Earth as a function of time for a northward (a, $\theta = 0°$) and westward landfall (b, $\theta = 270°$). The red dotted line marks t = 0, when the storm centre crosses the coastline. **c,d,** Wind speed contours at 10 m for the northward (c) and westward (d) cases at $t_a = -24$ h and $t_b = 24$ h. Grey dashed lines mark the coastline and black arrows indicate the direction of storm motion.

## Discussion

Our study provides, to our knowledge, the first estimates of how real landfalling tropical cyclones perturb Earth's rotation. Unlike earthquake-induced changes, which can persist long after the event, the effect of a landfalling tropical cyclone is expected to diminish as the atmospheric circulation relaxes towards its long-term mean state (typically weeks)[28,29]. Nevertheless, the storm-induced ΔLOD can rival that by a great earthquake.

Contrary to the braked-flywheel picture, a landfalling tropical cyclone can either lengthen or shorten the day, depending on its landfall direction. This does not violate angular-momentum conservation. The storm is an open system that also exchanges angular momentum with the surrounding atmosphere. The storm's decay, therefore, does not necessarily imply a gain of prograde angular momentum by the solid Earth. The angular momentum transferred to the solid Earth is instead set by the time-integrated axial torque exerted by surface stresses.

Climate change may further modulate storm-induced ΔLOD. Projections indicate an increase in the mean intensity of tropical cyclones and in the proportion reaching very high intensities[30]. Landfalling storms may also decay more slowly in a warmer climate[14]. Stronger winds and slower post-landfall decay would increase both the magnitude and the duration of the surface frictional torque, and therefore increase |ΔLOD|. Observations further show that tropical cyclones now reach their maximum intensity progressively closer to land[31], and some projections indicate a higher proportion of U.S. landfalling storms[32]. Assuming no systematic shift in landfall direction, these changes imply that storm-induced ΔLOD perturbations may become larger and more frequent in a warming climate.

# Methods

## Selecting landfalling tropical cyclones

We select representative landfalling tropical cyclones from the North Atlantic, where the HURDAT2 best-track record is particularly well constrained[17]. We consider only continental landfalls; landfalls on narrow peninsulas (for example, the Florida peninsula) or islands (for example, Cuba and the Bahamas) are excluded.

From the NOAA/NCEI ranking of the costliest U.S. tropical cyclones during 1983–2024[18], we select the ten highest-ranked storms that have at least four consecutive 6-hourly records over land before dissipation, re-emergence over the ocean, or extratropical transition. For storms with multiple qualifying continental landfalls, we retain the landfall with the highest intensity.

We additionally include Hurricanes Mitch (1998) and Isidore (2002), which made landfall while moving southward and eastward, respectively, thereby broadening the range of landfall directions over which the dependence of ΔLOD on storm motion can be examined.

## ERA5 Estimation of ΔLOD

The 6-hourly HURDAT2 storm-centre positions are linearly interpolated to an hourly time step[17]. ERA5 grid cells with land–sea mask $lsm \geq 0.5$ are classified as land. We consider only the torque arising from surface stress over land. (The torque exerted on the ocean is transmitted to the solid Earth much more slowly, on timescales much longer than the typical decay timescale of landfalling tropical cyclones[33].) ERA5 surface stress includes parameterised turbulent orographic form drag, which is not explicitly represented in our idealised CM1 simulations. The eastward surface stress is obtained from the hourly accumulated ERA5 eastward turbulent surface stress (*ewss*) as $\sigma_\lambda = ewss/t_{\text{step}}$ ($t_{\text{step}}$ = 3600s)[34]. The axial torque on the solid Earth is calculated as

$$\Gamma(t) = \iint_{A(t)} R\cos\varphi \, \sigma_\lambda(\lambda, \varphi, t) \, dA$$

where $A$(t) is the land area within 500 km of the storm centre, R is Earth's radius and $\varphi$ is latitude. The cumulative angular-momentum transfer is calculated as

$$\Delta L = \int_{t_0}^{t_1} \Gamma(t) \, dt$$

where $t_0$ is the time at which the 500-km disc first overlaps land and $t_1$ is the time at which the intensity of the storm weakens below 17.5 m $s^{-1}$ or the storm undergoes extratropical transition according to HURDAT2. The corresponding change in the length of day is

$$\Delta\mathrm{LOD} \approx -\frac{\mathrm{LOD}_0}{I_m \Omega_0} \Delta L$$

where Earth's mean angular velocity $\Omega_0 = 7.29 \times 10^{-5}$ $s^{-1}$ and $\mathrm{LOD}_0 = 86400$ s. We use the crust-mantle axial moment of inertia $I_m \approx 7.12 \times 10^{37}$ kg $m^2$.[35]

**Numerical Simulations**

We simulate landfalling tropical cyclones with Cloud Model 1 (CM1, v20.3)[22,23]; simulation parameters are listed in Extended Data Table 3. The domain is split into a land half and an ocean half. Over land, we set the surface drag coefficient to $C_D$ = 0.005 and the enthalpy exchange coefficient to $C_E = 0$; over ocean, the default surface fluxes and wind-dependent $C_D$ ranges from 0.001 to 0.0024[36,37,38]. After the storm develops over ocean for 8 days and reaches a mature state, a background steering flow is imposed to steer it towards land. Below 500 m, the zonal flow follows[26]

$$U = -\frac{u^*}{\kappa} \ln\left(\frac{Z}{Z_0}\right)$$

with friction velocity $u^* = 0.136$ m $s^{-1}$, von Kármán constant $\kappa = 0.4$, and roughness length $Z_0 = 0.0002$ m; above 500 m it is held constant at −5 m $s^{-1}$.

The Coriolis parameter is held constant at its value at 20°N. To represent different landfall directions, we rotate the simulated wind field through 360° at 3° intervals, yielding 120 orientations. For each orientation, the eastward surface stress is calculated as

$$\sigma_\lambda = \rho\, C_D\, u_{10}\, |V_{10}|$$

where $V_{10} = (u_{10}^2 + v_{10}^2)^{1/2}$ and $\rho$ is the air density at the lowest model level. The landfall point is fixed at 20°N, with grid-point latitude determined from meridional displacement from the landfall point. Axial torque and ΔLOD are then calculated as in the ERA5 analysis within the 500-km storm-following land mask, following the same procedure as in the ERA5 analysis. Integration ends when the 10-m maximum wind speed falls below tropical-storm intensity (17.5 m $s^{-1}$).

## Data availability

The ERA5 reanalysis data used in this study, including the hourly turbulent surface-stress fields and land–sea mask, are publicly available from the Copernicus Climate Data Store in the dataset “ERA5 hourly data on single levels from 1940 to present” (https://cds.climate.copernicus.eu/datasets/reanalysis-era5-single-levels). The Atlantic HURDAT2 best-track data used to determine storm-centre positions, intensities, landfall times and storm status are publicly available from the National Hurricane Center at https://www.aoml.noaa.gov/hrd/hurdat. The version of these data used in this study was downloaded on 22 January 2026. The earthquake-induced length-of-day changes used for comparison are compiled from previously published estimates. The values for individual earthquakes and their corresponding literature sources are provided in Extended Data Table 2. The CM1 model output analysed in this study is available for download as NetCDF file (.nc) without restrictions from: https://drive.google.com/drive/folders/1lhTO2prNoTLd6_fBaIdntCrW_Nu7zj8d.

## Code availability

The numerical simulations are performed using Cloud Model 1 (CM1), version 20.3, which is publicly available from the official CM1 website. The files required to reproduce the CM1 simulations in this study are available without restriction at https://drive.google.com/drive/folders/1lhTO2prNoTLd6_fBaIdntCrW_Nu7zj8d. The run directory contains the namelist and background-wind input files used for the simulations, whereas the src directory contains the three modified Fortran source files used to implement the half-ocean, half-land lower-boundary configuration.

## Author contributions

L.Li. conceived the study and directed the project. Y.Y. carried out the reanalysis-based estimates of LOD change, performed the numerical simulations, and analysed the data. L.Liu assisted with the set-up of the idealised ocean–land model in CM1 and with validation of the results. Y.L. contributed to the data analysis. L.Li and Q.W. supervised the work. Y.Y. and Z.Z. wrote the initial draft. Y.Y. and L.Li revised and finalised the manuscript. All authors discussed the results and commented on the manuscript.

### Funding statement

L. Li. acknowledges support from the National Natural Science Foundation of China through the Excellent Young Scientists Fund Program (Overseas) and the General Program (Grant No. 12572474), and from the Fundamental Research Funds for the Central Universities of China.

**Extended Data Table 1 | Representative tropical cyclones**

| Tropical cyclones | Year | Landfall category | ΔLOD (μs) |
|---|---|---|---|
| Katrina | 2005 | 3 | +2.29 |
| Harvey | 2017 | 3 | +1.87 |
| Ida | 2021 | 4 | +1.82 |
| Andrew | 1992 | 3 | +1.34 |
| Ike | 2008 | 2 | +2.62 |
| Ivan | 2004 | 3 | +2.57 |
| Michael | 2018 | 5 | +0.28 |
| Florence | 2018 | 1 | +3.05 |
| Rita | 2005 | 3 | +3.41 |
| Laura | 2020 | 4 | +1.36 |
| Mitch | 1998 | 1 | −3.43 |
| Isidore | 2002 | 3 | −1.26 |

**Extended Data Table 2 | Earthquakes used for comparison**

| Region | Year | Mw | ΔLOD (μs) |
|---|---|---|---|
| Chile[7] | 1960 | 9.5 | −8.40 |
| Alaska[7] | 1964 | 9.2 | +6.80 |
| Sumba[7] | 1977 | 8.3 | +0.33 |
| Chile[7] | 1985 | 8.0 | −0.10 |
| Mexico[7] | 1985 | 8.0 | −0.09 |
| Sumatra[11] | 2004 | 9.1 | −6.80 |
| Chile[11] | 2010 | 8.8 | −1.26 |
| Tōhoku[11] | 2011 | 9.1 | −1.80 |

**Extended Data Table 3 | Parameters for numerical simulations**

| Items | Setting |
|---|---|
| Domain dimensions | 6000 km (*x*) × 3000 km (*y*) × 25 km (*z*) |
| Spatial resolution | 567 (*x*) × 192 (*y*) × 59 (*z*) |
| Horizontal grid spacing | In the *x* direction, grid spacing is uniform at 8 km across the central 3560 km of the domain, then gradually stretches to 32 km towards both lateral boundaries. In the *y* direction, it is uniform at 8 km across the central 560 km, then stretches to 32 km towards both boundaries. |
| Vertical grid spacing | Vertical grid spacing is 50 m near the surface, stretched to 500 m at 5.5 km, and held uniform at 500 m from there to the 25 km model top. |
| Turbulence model | Simple PBL parameterisation with horizontal Smagorinsky mixing[22,23]. Vertical diffusion is treated implicitly. |
| Boundary conditions | Lateral: periodic.<br>Rayleigh damping above 20 km and within 100 km of the lateral boundaries.<br>Bottom: over ocean, $C_D$ varies with V ($C_D$ = 0.001 for $V \leq$ 5m $s^{-1}$; for $5 < V \leq 25$ m $s^{-1}$, $C_D$ linearly increases from 0.001 to 0.0024; for $V > 25$ m $s^{-1}$, $C_D$ = 0.0024)[36,37,38], and $C_E$ = 0.0012. Over land, $C_D$ = 0.005 and $C_E$ = 0. |
| Sea-surface temperature | 301 K |
| Initial condition | A weak vortex (15 m $s^{-1}$) |
| Coriolis force | $5.0 \times 10^{-5}$ $s^{-1}$ |
| Time stepping | Adaptive time stepping was enabled; the initial target large time step was 60 s, with the actual time step adjusted automatically to maintain numerical stability. |
| Sounding | "Dunion_MT sounding[39]", a benchmark sounding for tropical cyclone simulations in CM1 |

(x,y,z) are the axes in a Cartesian coordinate system with (x, y) spanning a horizontal plane, z is the vertical coordinate measured upward from the lower boundary.

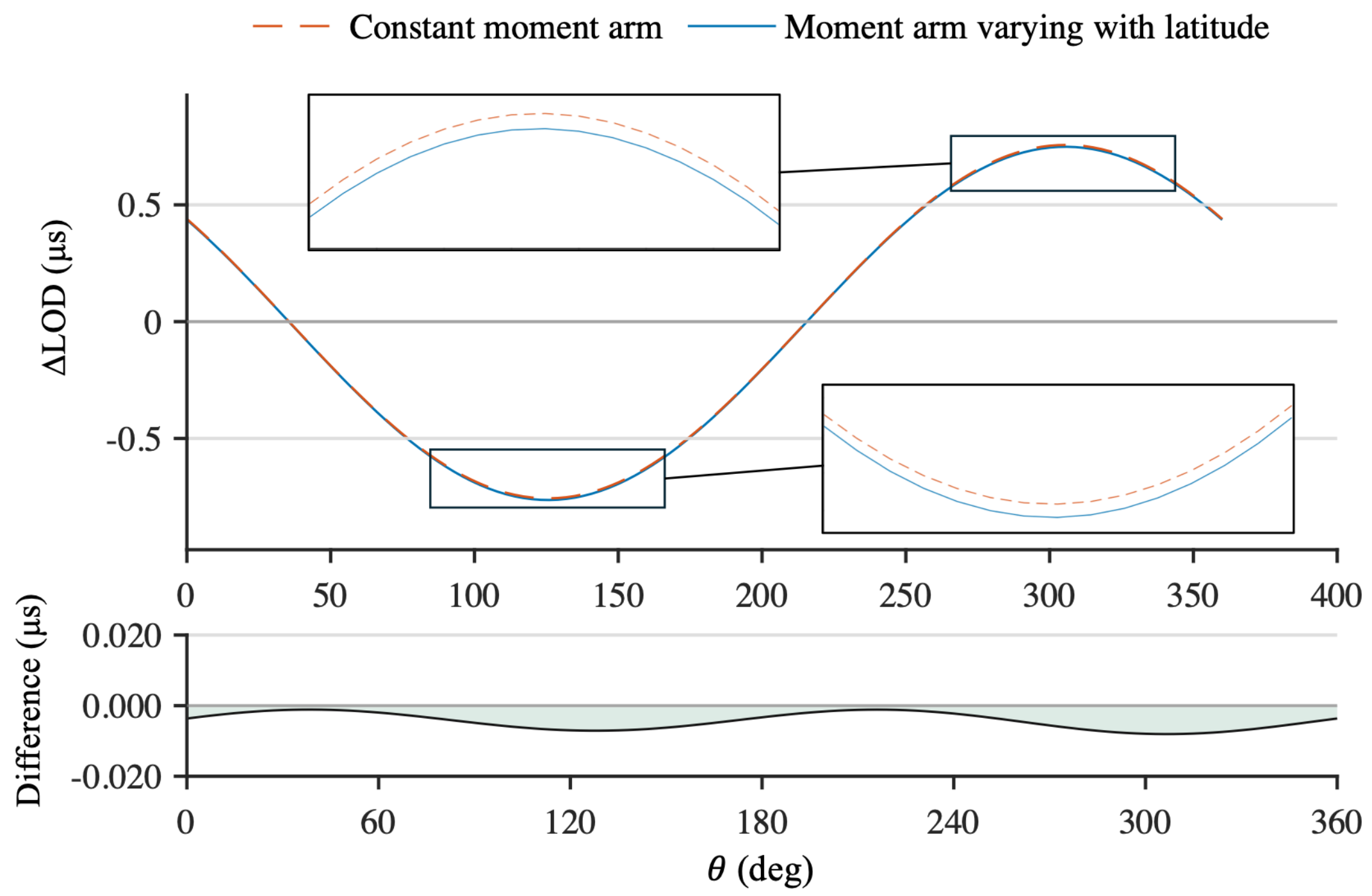

**Extended Data Figure 1 | The Rankine-vortex model reproduces the directional dependence of ΔLOD in CM1.**

**a**, ΔLOD as a function of landfall direction $\theta$ for Rankine vortex. The curve is nearly sinusoidal and qualitatively matches the form obtained with CM1 (Fig. 2). Two cases are shown: the moment arm held constant at 20° N (dashed orange), and the moment arm varying with latitude across the storm (solid blue). The two are nearly indistinguishable. The insets magnify the positive and negative extrema, where a small offset appears. **b**, Their difference (varying minus constant) stays within 0.01 μs, roughly two orders of magnitude smaller than the amplitude of ΔLOD.